\documentclass[preprint,5p]{elsarticle} 

\usepackage{lineno,hyperref}
\modulolinenumbers[5] 
\usepackage{amsmath,amsthm}
\usepackage{amsfonts,amssymb}
\usepackage{upgreek}
\usepackage{graphicx,color}
\usepackage{epstopdf} 
\usepackage[caption=false]{subfig}
\usepackage{multirow}
\usepackage{csquotes}

\biboptions{numbers,sort&compress} 

\begin{document}

\begin{frontmatter}

\title{Numerical benchmark of transient pressure-driven metallic melt flows}

\author[kth]{L. Vignitchouk\texorpdfstring{\corref{cor1}}{}}
\ead{ladislas.vignitchouk@ee.kth.se}
\author[pppl]{A. Khodak}
\author[kth]{S. Ratynskaia}
\author[pppl]{I. D. Kaganovich}

\cortext[cor1]{Corresponding author}
\address[kth]{Space and Plasma Physics, KTH Royal Institute of Technology, SE-100 44 Stockholm, Sweden}
\address[pppl]{Princeton Plasma Physics Laboratory, Princeton NJ 08543, USA}

\begin{abstract}
Fluid dynamics simulations of melting and crater formation at the surface of a copper cathode exposed to high plasma heat fluxes and pressure gradients are presented. The predicted deformations of the free surface and the temperature evolution inside the metal are benchmarked against previously published simulations. Despite the physical model being entirely hydrodynamic and ignoring a variety of plasma-surface interaction processes, the results are also shown to be remarkably consistent with the predictions of more advanced models, as well as experimental data. This provides a sound basis for future applications of similar models to studies of transient surface melting and droplet ejection from metallic plasma-facing components after disruptions.
\end{abstract}


\end{frontmatter}


\section{Introduction}

Transient surface melting can occur on metallic plasma-facing components (PFC) as a consequence of off-normal or pulsed energy loads, such as those associated with unipolar arcs, large edge-localized modes, and disruptions. If heat diffusion into the solid bulk is slow enough, the melt layer can flow under the action of plasma-induced forces before re-solidifying, which may lead to significant modifications of the surface morphology, thereby degrading the PFC power-handling capabilities~\cite{Pitts2017}. In some cases, part of the molten metal is ejected as droplets, which later solidify into dust particles. Signs of droplet production during transient melt events have in fact been observed on tungsten PFCs subject to arcing in ASDEX Upgrade~\cite{Rohde2016}, and on beryllium dump plates after disruptions in JET~\cite{Jepu2019}. In ITER, beryllium splashing during disruptions is expected to be a major contributor to the in-vessel dust inventory, which has to be maintained below certain limits~\cite{Shimada2013,Lehnen2015,Pitts2015}. While the conversion of these droplets into dust particles and their migration inside the vacuum vessel have been studied~\cite{Vignitchouk2018_2,Vignitchouk2019}, no complete prediction can be made without sufficient knowledge of the main characteristics of the droplet source, namely its magnitude as well as typical size and velocity distributions.

Experimental data on transient PFC melting in contemporary devices and the resulting dust production mainly stems from tokamak wall surveys after entire campaigns~\cite{Rohde2011,Rohde2013,Rudakov2013,Balden2014,Flanagan2015,BaronWiechec2015,Rohde2016,Bykov2017,Jepu2019} and from controlled exposures of specifically designed components to a few discharges~\cite{Coenen2015,Krieger2018}. In these studies, the measurements relate to the damaged surface after melting and re-solidification have occurred, yielding only partial information on the characteristics of the melt flow. Linear plasma generators offer more experimental flexibility~\cite{Bardin2015,Klimov2017}, allowing in some cases for direct measurements of the droplet velocity distributions~\cite{Bazylev2009_2,DeTemmerman2013,Siemroth2019,Makhlai2020}. However, these results are difficult to extrapolate to fusion conditions due to potentially major differences in materials or plasma environments.

On the theoretical front, the problem can be postulated as a magneto-hydrodynamic free-surface flow coupled with heat transfer and phase transitions. Designing self-consistent models or simulations of such systems is very challenging and computationally expensive, especially in cases where the flow is governed by multiple spatial and temporal scales. As a result, compromises must be made depending on which results are sought. For example, if the focus is set on the macroscopic flow features and large-scale melt displacements, simplifications such as the shallow water approximation~\cite{Vreugdenhil1994} have been shown to reproduce the experimental data accurately and at reasonable computational costs~\cite{Bazylev2002,Bazylev2013,Thoren2018_2,MEMOS_NF,MEMOS_PPCF}. However, the shallow water equations are not suited to study melt ejection, flow instabilities or strong deformations of the free surface. These processes require the solution of the full set of Navier-Stokes equations in a multi-phase framework.

Such numerical studies are rare in fusion-relevant conditions, consisting mostly of simulations of shear-driven flows in prescribed liquid pools~\cite{Miloshevsky2010,Miloshevsky2011,Miloshevsky2014} and simulations of pressure-driven flows intended to mimic the initial stage of unipolar arcs~\cite{Kaufmann2019}. Nevertheless, major similarities exist with problems that arise in laboratory and industrial arc discharges, for which the available literature is more extensive. In these settings, melting and free-surface dynamics play a central role in a variety of phenomena such as droplet transfer in welding~\cite{Boselli2012,Hertel2013,Hertel2016} and jet formation in the vicinity of cathode spots~\cite{Mesyats2015,Mesyats2017,Cunha2017,Kaufmann2017,Kaufmann2018,Cunha2019,Zhang2019}. More general plasma-surface interaction processes have also been shown to govern the properties of atmospheric pressure arcs~\cite{Khrabry2018_1,Khrabry2018_2,Khrabry2019}. These studies often make use of comprehensive multiphysics or computational fluid dynamics software such as COMSOL Multiphysics, ANSYS CFX, ANSYS Fluent or OpenFOAM, which give access to a wide range of numerical solvers while still allowing for some degree of customization. However, the large number of physical and numerical parameters make benchmarking against known solutions and validation against experiments of utmost importance. In the absence of detailed experimental results for fusion problems, one must resort to cross-benchmarks with published simulations of similar problems.

This paper presents a benchmark of liquid flows at the surface of a copper (Cu) cathode exposed to prescribed heat fluxes and pressure gradients intended to emulate those from the arc plasma in the vicinity of a cathode spot. The selected benchmark case is that described in~\cite{Mesyats2015} which, while not being the most advanced model in the literature~\cite{Mesyats2017,Cunha2017,Kaufmann2017,Kaufmann2018,Cunha2019,Zhang2019}, has the advantage to be self-contained and transparent enough for a reproduction study. First, we repeat the simulations of~\cite{Mesyats2015} and compare our results obtained with a customized set-up in ANSYS Fluent to those previously published as far as cathode deformations are concerned. We then extend some of our simulations until droplet detachment occurs and show that, despite its relative simplicity, the model is able to produce qualitatively and quantitatively similar results to its upgraded counterparts.

\section{Numerical set-up}

\subsection{Model description}

\begin{figure}
\centering
\includegraphics[width=0.9\columnwidth]{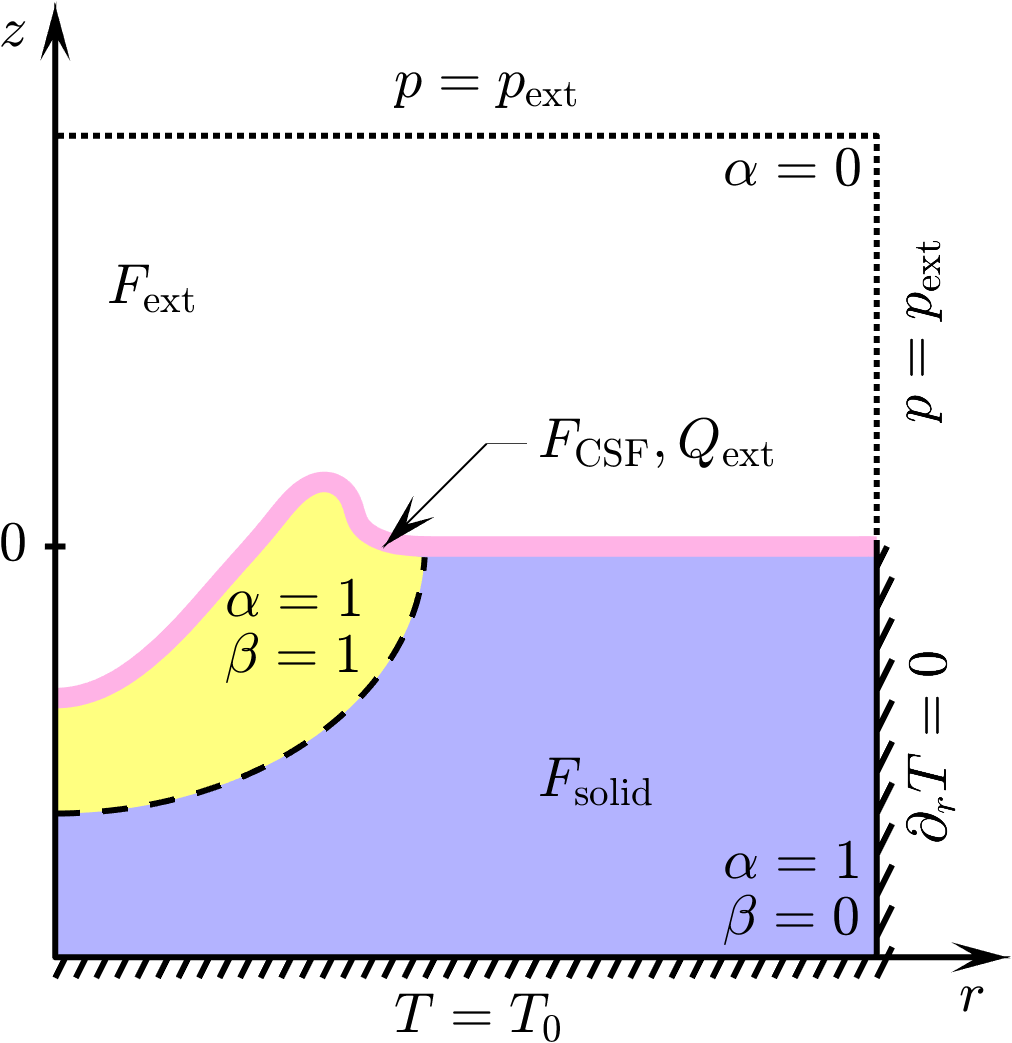}
\caption{Computational domain used in the simulations. The boundary conditions, source terms and phase fractions are indicated.}
\label{fig:domain}
\end{figure}

The physical model and simulation geometry are chosen to be as close as possible to those described in~\cite{Mesyats2015}. Fluent simulations are run on a two-dimensional axisymmetric domain whose section in the meridian half-plane is an $8\times 8~\mu\text{m}^2$ square, as shown in Fig.~\ref{fig:domain}. Initially, the lower half ($z<0$) of the domain is occupied by the solid Cu cathode at 300~K. The plasma heat flux and pressure are imposed respectively as $q_\text{ext}\left(r,t\right) = q_0f_r(r)f_t(t)$ and $p_\text{ext}\left(r,t\right) = p_0f_r(r)f_t(t)$, where

\begin{equation}
f_r(r) = \frac{1}{1+r^2/r_0^2}\,,
\end{equation}

\begin{equation}
f_t(t) = \begin{cases}
1 & \text{if } t\le t_0\,, \\
\exp\left(-\left(\frac{t-t_0}{\tau}\right)^2\right) & \text{if } t\ge t_0\,.
\end{cases}
\end{equation}

\noindent The simulations presented in this paper use $q_0=1.5\times 10^{12}~\text{W}~\text{m}^{-2}$ for the central plasma heat flux, $r_0=1~\mu\text{m}$ for the characteristic radial decay length, $t_0=30$~ns for the plasma duration and $\tau=1$~ns for the characteristic plasma decay time. The central plasma pressure $p_0$ is varied between $0.3\times 10^8$~Pa and $4\times 10^8$~Pa. Note that $p_\text{ext}$ and $q_\text{ext}$ is assumed to be the only momentum and heat exchange channels in the model, which means in particular that other potentially relevant physical processes such as thermal radiation, Joule heating or vaporization are ignored.

As time progresses, the plasma heat flux melts the central part of the Cu volume and the resulting liquid pool starts accelerating outwards under the action of the external pressure gradient, which leads to the formation of a crater and a rim. At low values of $p_0$ and $q_0$, the rims cool down via heat diffusion into the cathode bulk and quickly re-solidify once the plasma vanishes. If larger heat fluxes and pressure gradients are applied, the rims can remain liquid longer and extend into jets, possibly resulting in the ejection of Cu droplets.

\subsection{Equations, boundary conditions and material properties}\label{sec:modeleq}

The main equations of the model are the incompressible Navier-Stokes and the heat convection-diffusion equations:

\begin{equation}
\nabla \cdot \boldsymbol{v} = 0 \,,
\end{equation}

\begin{equation}\label{eq:continuity}
\frac{\partial \rho}{\partial t}+\nabla\cdot\left(\rho\boldsymbol{v}\right) = 0 \,,
\end{equation}

\begin{equation}\label{eq:momentum}
\frac{\partial}{\partial t}\left(\rho\boldsymbol{v}\right)+\nabla\cdot\left(\rho\boldsymbol{v}\boldsymbol{v}\right) = -\nabla{p} + \nabla\cdot\left[\mu\left(\nabla\boldsymbol{v}+\nabla\boldsymbol{v}^\text{T}\right)\right] + \boldsymbol{F}\,,
\end{equation}

\begin{equation}\label{eq:heat}
\frac{\partial}{\partial t}\left(\rho h\right)+\nabla\cdot\left(\rho h\boldsymbol{v}\right) = \nabla\cdot\left(\lambda\nabla T\right) + Q\,,
\end{equation}

\noindent where $\rho$ is the mass density, $\boldsymbol{v}$ the velocity, $p$ the pressure, $\mu$ the dynamic viscosity, $h$ the specific enthalpy, $T$ the temperature and $\lambda$ the thermal conductivity. The role of the external force density $\boldsymbol{F}$ and the external volumetric power $Q$ is detailed in the following paragraphs.

These equations are solved on a fixed cartesian mesh using the PISO pressure-velocity coupling algorithm~\cite{Issa1986} and the volume of fluid (VOF) method~\cite{Hirt1981} to track the position of the free surface between the metal and the plasma. In a VOF setting, Eq.~(\ref{eq:continuity}) is replaced by

\begin{equation}
\frac{\partial \alpha}{\partial t}+\nabla\cdot\left(\alpha\boldsymbol{v}\right) = 0 \,,
\end{equation}

\noindent where $\alpha$ is the metal volume fraction, which evaluates to $0$ in the plasma and $1$ in the metal, with a sharp transition at the free surface. No formal distinction is made between the phases: all phases share the same velocity, pressure and enthalpy fields; the computational domain is effectively filled with a single fluid whose material properties are obtained as weighted sums, such as

 \begin{equation}
\rho = \alpha\rho_\text{m} + \left(1-\alpha\right)\rho_\text{p} \,,
\end{equation}

\noindent where the subscripts `m' and `p' respectively refer to the metal and the plasma.

The boundary conditions imposed on the simulation domain are detailed in Fig.~\ref{fig:domain}: the bottom boundary is an isothermal no-slip wall at $300$~K, the bottom half of the outer boundary is an adiabatic no-slip wall, and the remaining boundary is an opening with imposed pressure $p_\text{ext}$. Since the free surface is not a computational boundary in the VOF formalism, boundary conditions cannot be imposed directly on it and have to be reformulated into volumetric source terms in Eqs.~(\ref{eq:momentum}-\ref{eq:heat}). Here, we use

\begin{equation}
\boldsymbol{F}_\text{ext} = \left(1-\alpha\right)\nabla p_\text{ext} \,,
\end{equation}

\begin{equation}\label{eq:interfaceheat}
Q_\text{ext} = 2\left|\nabla\alpha\right| q_\text{ext} \,,
\end{equation}

\noindent which serve to impose $p=p_\text{ext}$ in the plasma and produce a localized heat source at the interface. The factor $2$ in Eq.~(\ref{eq:interfaceheat}) derives from the assumption that the free surface is equally shared between the metal and the plasma, so that each phase receives half of the energy source. Since the physics of the flow in the plasma are of no interest in the present problem, the plasma properties used in the simulations do not have to match those expected of a real plasma. Rather, they can be chosen to ensure that the plasma does not affect the metal flow in any manner other than enforcing the desired boundary conditions at the free surface --- in particular, one must make sure that the plasma does not resist the motion of the metal and that heat cannot diffuse across the free surface. To this end, the plasma is set to have very low density, viscosity and conductivity,

\begin{equation}
\rho_\text{p}/\rho_\text{m}\sim \mu_\text{p}/\mu_\text{m} \sim \lambda_\text{p}/\lambda_\text{m} \sim 10^{-6} \,,
\end{equation}

\noindent while its volumetric heat capacity is set to be similar to that of the metal so as to avoid strong temperature gradients at the interface. The physical properties of Cu are taken from the same references as in~\cite{Mesyats2015}: $\rho_\text{m}=8000~\text{kg}~\text{m}^{-3}$, $\mu_\text{m}=3.3\times 10^{-3}~\text{N}~\text{s}~\text{m}^{-2}$, $\lambda_\text{m}$ and $c_\text{p,m}=\text{d}h_\text{m}/\text{d}T$ from~\cite{Zinoviev1989}. The values of the specific heat capacity $c_\text{p,m}$ of Cu below the melting point ($T_\text{melt}=1358$~K) are corrected to ensure that the real volumetric heat capacity $\rho_\text{m}c_\text{p,m}$ of solid Cu is recovered, in spite of $\rho_\text{m}$ being the density of liquid Cu.

Surface tension effects and solid-like behavior below $T_\text{melt}$ are also modelled by extra source terms in Eq.~(\ref{eq:momentum}), which are natively available in Fluent. The former is incorporated via the continuum surface force model~\cite{Brackbill1992}, wherein the interfacial force is written as

\begin{equation}
\boldsymbol{F}_\text{CSF} = \gamma \kappa \nabla \alpha \,,
\end{equation}

\noindent where $\gamma=1.12~\text{N}~\text{m}^{-1}$~\cite{Mesyats2015} is the surface tension and $\kappa$ is the algebraic curvature of the free surface, estimated from the spatial derivatives of $\alpha$. Solidification and melting are implemented following the enthalpy-porosity method~\cite{Brent1988}, wherein the molten fraction $\beta$ is defined as

\begin{equation}
\beta = \begin{cases}
0 & \text{if } T \le T_\text{melt}\,, \\
1 & \text{if } T \ge T_\text{melt}+\Delta T_\text{melt}\,, \\
\displaystyle\frac{T-T_\text{melt}}{\Delta T_\text{melt}} & \text{if } T_\text{melt} \le T \le T_\text{melt}+\Delta T_\text{melt}\,,
\end{cases}
\end{equation}

\noindent where $\Delta T_\text{melt}$ is a small temperature region over which the solid-liquid transition occurs (here, $\Delta T_\text{melt}=1$~K is used). The solid region of the metal is then frozen in place by a spring-like force

\begin{equation}
\boldsymbol{F}_\text{solid} = -\alpha\frac{\left(1-\beta\right)^2}{\beta^3+\epsilon}A\boldsymbol{v}\,,
\end{equation}

\noindent where $A=10^{12}~\text{N}~\text{s}~\text{m}^{-4}$ is the so-called mushy zone constant and $\epsilon=10^{-3}$ is a small dimensionless number meant to avoid division by zero. The latent heat of fusion $\Delta h_\text{melt}=213~\text{kJ}~\text{kg}^{-1}$~\cite{Mesyats2017} is accounted for in the heat equation by writing

\begin{equation}
h_\text{m} = h_\text{m}^0 + \int_{T_0}^T c_\text{p,m}\left(T^\prime\right)\text{d}T^\prime + \beta\Delta h_\text{melt}\,,
\end{equation}

\noindent where $h_\text{m}^0$ and $T_0$ are the reference enthalpy and temperature (here $h_\text{m}^0=0$ and $T_0=300$~K).

The complete form of the source terms of Eqs.~(\ref{eq:momentum}-\ref{eq:heat}) used in the simulations described in this paper is

\begin{equation}
Q = Q_\text{ext}\,,
\end{equation}

\begin{equation}
\boldsymbol{F}=\boldsymbol{F}_\text{ext}+\boldsymbol{F}_\text{CSF}+\boldsymbol{F}_\text{solid}\,.
\end{equation}

\noindent Interface processes that are neglected here, such as vaporization and thermal radiation, can in principle be implemented in a similar fashion.

\begin{figure}
\centering
\includegraphics[width=0.87\columnwidth]{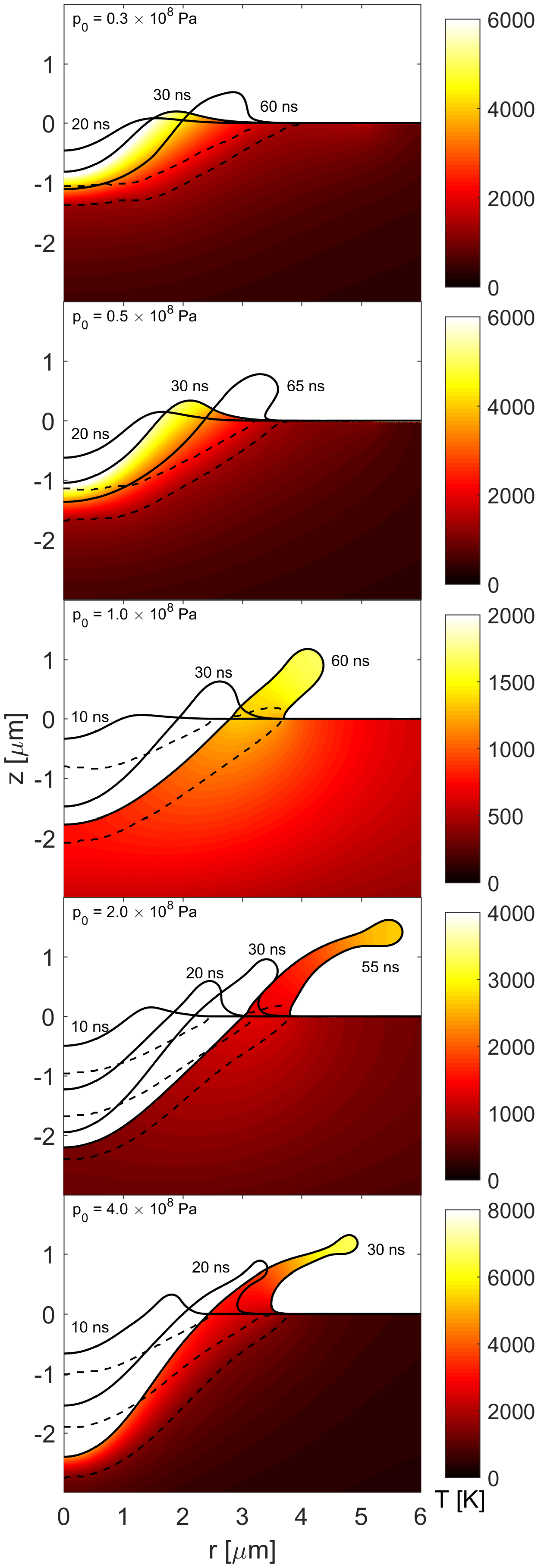}
\caption{Simulated temperature response for varying values of the central pressure. The free surface (solid lines) and melt front (dashed lines) are indicated at various times, similarly to Fig.~2 in~\cite{Mesyats2015}.}
\label{fig:benchmark}
\end{figure}

\section{Results and discussion}

\subsection{Comparison with previous simulations}

The simulation results concerning the Cu cathode deformations and temperature evolution are plotted in Fig.~\ref{fig:benchmark} for various values of the central pressure $p_0$. The shape of the free surface is shown at several instants, identical to those used in Fig.~2 of~\cite{Mesyats2015} to facilitate comparison. The characteristic crater and rim structure is very well reproduced by Fluent simulations, as well as the formation of jets when $p_0$ exceeds $10^8$~Pa.

Some discrepancies with the results of~\cite{Mesyats2015} can be observed in the temperature values: Fluent simulations seemingly lead to higher temperatures in the two upper panels (i.e. near the center of the cathode, or at early times before the plasma vanishes), but lower temperatures in the three lower panels (i.e. in near the jet head, or at times after the plasma vanishes). These might be due to differences in how material properties are extrapolated at high temperatures --- the data of~\cite{Zinoviev1989} stops at $1600$~K but the Cu temperature in the simulations can reach above $7000$~K. Another possible cause might be linked to the form we assumed for the volumetric heat source term in Eq.~(\ref{eq:interfaceheat}).

\subsection{Droplet ejection}

\begin{figure}
\centering
\includegraphics[width=\columnwidth]{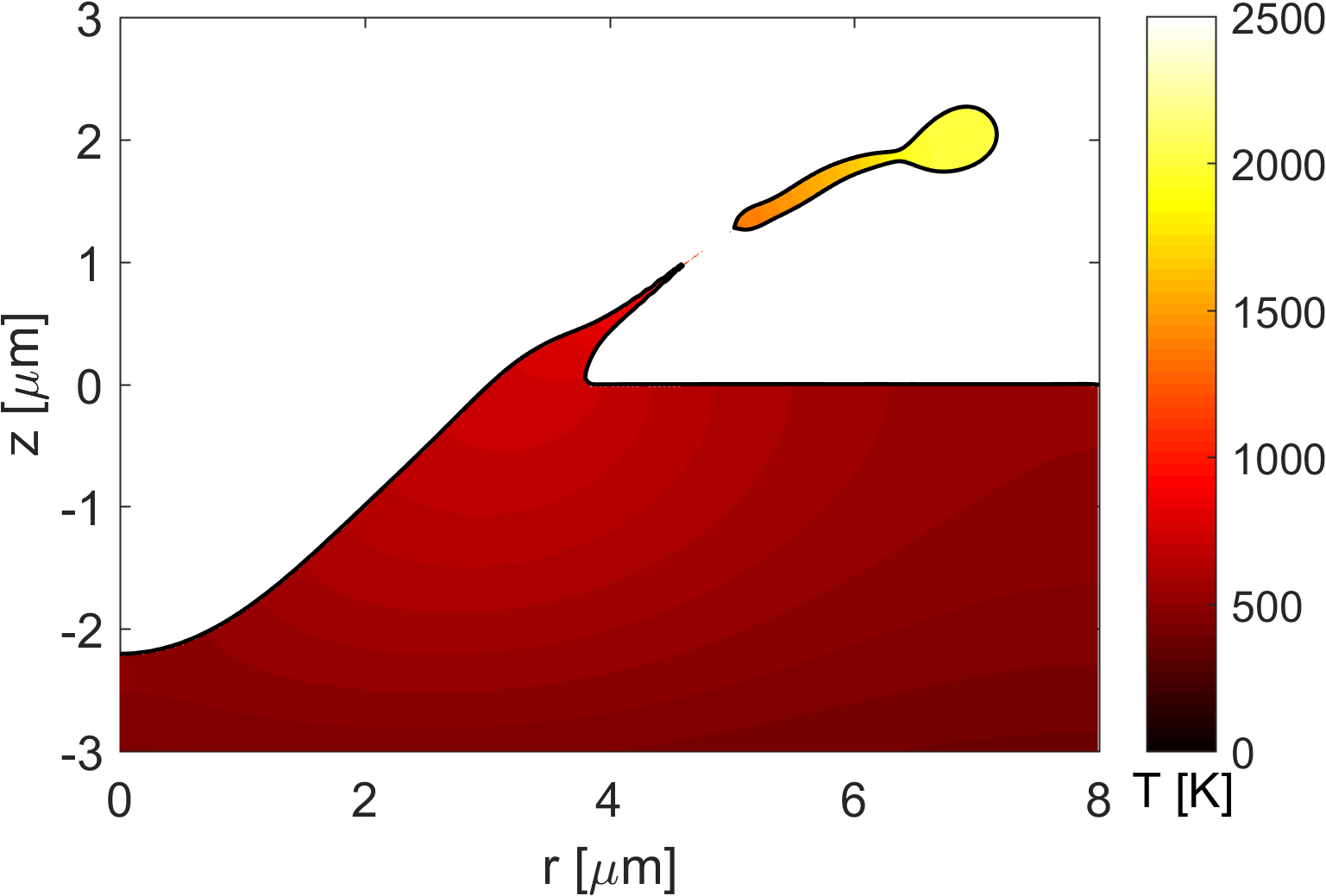}
\caption{Simulated temperature response after $80$~ns, with $p_0=2\times 10^8$~Pa. The crater and rims have entirely re-solidified.}
\label{fig:droplet}
\end{figure}

The simulations presented in~\cite{Mesyats2015} are interrupted before the free surface breaks up on the basis that the predicted jet temperatures become unphysically high. Nevertheless, useful information can be gained by extending the simulation time and observing the characteristics of the melt ejection process. Fig.~\ref{fig:droplet} shows an example of such a continued calculation for $p_0=2\times 10^8$~Pa. It can be seen that detachment occurs due to the solidification front progressing into the jet and canceling its velocity, while the liquid head retains significant momentum due to its inertia. The resulting tension results in a thinning of the middle of the jet, ultimately leading to separation. The liquid ejecta retain a strongly elongated shape, which may lead to further break-up into smaller droplets due to Plateau-Rayleigh instabilities, while the re-solidified part of the jet features a very sharp point due to the thinning process described above.

The geometrical characteristics of the ejected droplet and the frozen remains of the jet are very similar to those simulated in~\cite{Kaufmann2017} (see e.g. Fig.~2 and 3 therein) under comparable plasma conditions. This is particularly noteworthy because the physical model implemented in~\cite{Kaufmann2017} accounts for plasma-surface interactions which have been discarded in this paper, such as the contribution of the ionized metal vapor and field electron emission on the heat fluxes at the free surface. This indicates that the simplified model used here is able to produce predictions with good quantitative accuracy. Moreover, the ejected droplet shown in Fig.~\ref{fig:droplet} is found to have an equivalent diameter of $\sim 0.8~\mu$m and a velocity of $\sim 65~\text{m}~\text{s}^{-1}$ forming an angle of $\sim 30^{\circ}$ with respect to the unperturbed cathode surface. These values are consistent with those observed experimentally for Cu droplets ejected after exposure to a pulsed vacuum arc~\cite{Siemroth2019}.

\section{Conclusions and outlook}

Free-surface simulations of transient melt events have been carried out with a customized set-up in ANSYS Fluent, accounting for the coupling between fluid flow, heat transfer and phase transitions under prescribed heat and pressure loads. The results have been benchmarked against published simulations of cathode spots in vacuum arcs~\cite{Mesyats2015} and quantitative agreement has been shown with numerical and experimental data of droplet ejection under similar conditions~\cite{Kaufmann2017,Siemroth2019}.

This benchmark case is to serve as a basis for future upgrades of the underlying physical model, aimed at simulations of fusion-relevant scenarios, such as unipolar arcing on tungsten components and transient beryllium melting during disruptions. Said upgrades concern in particular the inclusion of surface cooling due to vaporization, electron emission and thermal radiation~\cite{MEMOS_NF,MEMOS_PPCF,Vignitchouk2014,Vignitchouk2018_2}, as well as the Lorentz force due to thermionic or disruption currents~\cite{Coenen2015,Krieger2018,Thoren2018_1,Thoren2018_2,Pitts2015,Matthews2016}. The latter may also require coupling with electromagnetic field equations, typically in the magnetostatic limit~\cite{Thoren2018_1,MEMOS_NF}.

It should however be emphasized that the computational cost of full Navier-Stokes models such as the one presented here are expected to be too high for accurate three-dimensional simulations of the entire regions impacted by transient melting. This is due to the multi-scale nature of such flows, for which an illustrative example is provided by the observations of molten beryllium on upper dump plates in JET~\cite{Jepu2019}. In this case, the melt pool is shown to have a typical depth of the order of $100~\mu$m, while its extent ranges between $1$~cm and $10$~cm~\cite{MEMOS_NF} and some small-scale features of the flow require a resolution of $10~\mu$m or below. These scale separation arguments are also expected to hold for ITER, where millimeter-deep pools covering several tens of centimeters are predicted~\cite{Coburn2020}. Therefore, a combined modelling approach is necessary, wherein the macroscopic characteristics of the flow calculated by cost-efficient codes~\cite{MEMOS_NF} are fed into more detailed simulations of restricted spatial domains.

\section*{Acknowledgments}

This work has been carried out within the framework of the EUROfusion Consortium (WPPFC) and has received funding from the Euratom research and training programme 2014--2018 and 2019--2020 under grant agreement No 633053. The views and opinions expressed herein do not necessarily reflect those of the European Commission. S. R. acknowledges the financial support of the Swedish Research Council under grant No 2018--05273.


\bibliography{biblio}

\end{document}